\documentclass[twocolumn,aps,prl,showpacs,superscriptaddress]{revtex4}
\usepackage[dvips]{graphicx}
\usepackage{bm}

\begin{document}
\title{
The Polyakov line, the Nishimori line and polymer networks standing a
smectic liquid crystal}

\author{L. V.~Elnikova}
\affiliation{
 A. I. Alikhanov Institute for Theoretical and Experimental Physics, \\
Bolshaya Cheremushkinskaya street 25, Moscow 117218, Russian Federation}

\date{\today}

\begin{abstract}
We once more specify an universality of a phase transition from smectic-A to nematic phase in
the crosslinked polymer network with a smectic liquid crystal, basing on the X-ray and $^1H$-NMR
spectroscopy experiments.
Comparing the superconducting
3D $XY$ gauge theory and typical spin-glass models for such systems, it is possible to clarify a type of a phase transition, which might be the reason of percolation.
\end{abstract}

\maketitle

There are well know encountering problems in soft materials,
polymers and liquid crystals, when at a quantum phase transitions, a system reveals
percolation properties \cite{Ostr2004, polym, polym2}. The presence of observations the phase
transition from a smectic-A ($SmA$) to a nematic ($N$) phase in
crosslinked elastomers by X-ray \cite{Ostr2004} and $^1$H-NMR
spectroscopy \cite{Sotta1}, paves the way for speculations on the
percolation threshold predictions and measurements.

This report is devoted to the analysis of percolation at the observations cited above, in which a first riddle is
formulated as follows. During which of phase transitions
(thermal or quantum, i. e. caused by an increase of a crosslink density) percolation is happened?

Elastomers are polymer networks composed by crosslinked polymer
chains. They consist nodes and links obeying ascertained rules of
self organization \cite{Grosb_UFN, deGennes_book}.

So, polymeric random network systems involve different types of
ordering, orientational and translational, caused by topological
defects (e. g. by dislocations, vortexes, and so on) \cite{Ostr2004,
7, 8}. In such a systems, an increase of a crosslink density may carry
out to percolation \cite{deGennes_book, Broderix_1997, Khokhlov_book}.

For mere polymers, the percolation problem is considered in the light
of analogy of dimers with connective conductors
\cite{Broderix_1997}, \cite{dG_76} --\cite{Broderix}, revealing the fractal dimensionality of a system.

For a polymer networks, there are applicable spin glass models,
resolving with replica methods (see the review \cite{0438}), the
cradles of which are sufficiently ancient \cite{Derrida,
Parisi_Virasoro_book}.

Here we face with a more complicated combination of such fluids,
namely, the smectic bulk phase and elastomer network with
their crosslinks (saying in a sketch of the aerogels description
\cite{Leo} and nomenclature accepted there and references therein).

In a smectic, the linear nontrivial disclinations ($3D$ vortexes or
dislocation loops \cite{Polyakov}) may appear \cite{Leo, Dasgupta},
as well as a point monopole-type defects, suspecting the system are
consisting monomers and dimers particles, corresponding to the
liquid crystal molecules and polymeric fragments respectively.
Percolation associates with condensation of vortexes in the high
temperature phase, in the deconfinement phase ($\beta<\beta_c$,
where temperature denotes as $\beta=\frac{1}{k_BT}$, $k_B$ is the
Bolzmann's constant, and $T$ is an absolute temperature, $\beta_c$
labels a critical temperature)
\cite{Polikarp49}. Fractal-dimension domains of this phase
are existing on a
non-integer lattice \cite{Polikarp}, where is the Polyakov line
plays role of the order parameter \cite{Diak}.

For smectic-A liquid crystals surviving a tilt second order transition to a nematic
phase, Dasgupta proposed the loop inverted analog of the superconductive $XY$
model \cite{Dasgupta}. The action of this model is
equivalent of the Villain's one \cite{Villain}, if in the $XY$ model,
the de Gennes's coefficient $K_1=0$ (what means an
anisotropic case) \cite{Dasgupta, Nelson1981, Dasgupta_perc}.

Vortexes in the $XY$ 3D model form loops (in the $XY$ 4D model of compact
electrodynamics \cite{Polikarp49} they corresponds to strings). The $XY$
model is a quantized variant of a scalar field theory $\Phi$ with
broken global $U(1)$ symmetry. The potential of this theory is
$\lambda(|\Phi|^2-\Phi_0^2)^2$, at $\lambda \longrightarrow \infty$,
the radial term of $|\Phi|$ is freezing, and the residuary dynamical
variable $\varphi$ is compact ($\Phi=|\Phi_0| \exp(i \varphi)$)
\cite{Polikarp}.
A vortex, as well as a dislocation with its similar properties, is
non-trivial classical minimum of the action \cite{Polyakov}.

If a crosslinked elastomer system with smectic liquid crystal is considered in the 3D superconductivity
class \cite{Leo, sc_vort} with a scalar gauge field $k=0$, a
dimensionality of defects (vortices) then is $D-k-2=1$,
\cite{Polikarp}. As it was shown numerically \cite{Qian} for a bulk nematic, due to the coupling between smectic and nematic order
parameters and director fluctuations, the $SmA-N$
transition deviates from 3D $XY$ universality to a nonuniversal crossover
value between 3D $XY$ and tricritical behavior with an effective
heat capacity critical exponent $\alpha$ ranging between 0.26 and
0.31 confinements. In 3D, similar percolation holds below a critical
temperature, $\beta < \beta_c$, $\beta_c \approx 0.4542$
\cite{Polikarp}. The calculation procedure for $D_f$ is contained in \cite{Polikarp}; the gauge field $A$ \cite{Polikarp49},
however, makes difficult a quantitative characterization of the order
parameter in view of an ambiguity in $A$
\cite{Dasgupta}.

The order parameter $\langle Tr L \rangle$ in terms of the Polyakov line \cite{Diak}
\begin{equation}
L(x)={\mathcal{P}}\exp(i\int_0^{1/T}dx_4A_4(x_4,x))
\end{equation}
(where $\mathcal{P}$ denotes path ordering, $x_4$ and $A_4$ are a time coordinate component and a time component of the
 Yang-Mills field for 3D, in our case these are a temperature components) is zero in the confined phase below the critical 
temperature, and, otherwise, its a nonzero value denotes the broken symmetry.
But in contrast to usual Polyakov loops, our system is uncentered (such a $SmA-N$ transition
carries out 
from $T(2)\times D_{\infty h}$ to $D_{\infty h}\times T(3)$) \cite{Kats-Leb}. However, amid this transition, 
an exact sequence with $Z(N)$ should exist to provide $U(1)$ in the limit case $N \rightarrow \infty$. 
For instance, $Z(N)$ has to be a group of crosslinks of a dual lattice. 

The Polyakov line is not invariant under spatial gauge transformations. A periodical gauge transformation reads here as follows:
\begin{equation}
A_4(x)\rightarrow U(x)A_4(x)U(x)^{\dag},
\end{equation}
\begin{equation}
A_i(x)\rightarrow U(x)A_iU(x)^{\dag}+iU(x)\partial_iU(x)^{\dag}.
\end{equation}
And some a gauge field $\textsl{A}_{\mu}$ means a background field \cite{Diak, Diak13}; it has gauged in the same way:
\begin{equation}
A_{\mu}\rightarrow SA_{\mu}S^{\dag}+iS\partial_{\mu}S^{\dag}.
\end{equation}
As it was shown, the general $SU(N)$ gauge group has the explicitly symmetric $Z(N)$ resulting action. 
So, for $Z(3)$, the maxima of the potential energy 
of the system corespond to the Polyakov line $L_{max}=diag(1, e^{2\pi i/3}, e^{-2\pi i/3})$, Tr$L_{max}$=0, etc. \cite{Diak}.

Alternative models of the random pinning potential \cite{Ostr2004}, with the
appointed role of topological defects in polymers are
treatable as spin glassy approaches \cite{Martinez}, which may be resolvable by the replica tricks.
Also, a  random graph method \cite{Kosterlitz} is applied, which is allows us to take into account both a 
long-wavelength and short-wavelength fluctuations. For $n>1$ component
spin system, a modification of the Polyakov method is used.
These well developed ideas are not confirmed in the concrete case of the order parameter of polymers, 
complicated by smectic states \cite{Horacio}.

Progressing in the problem of $SmA-N$ phase transitions, several 2D cases of quantum smectics, 
breaking their symmetry into a nematic phase, were discussed in \cite{Cvetkovic}.
Properties of 2+1D dislocation loops there were corresponded to the quantum double
Hopf symmetry.

On the other hand, the spin-glass theory of polymers via its accessible results of a replica approximation gives a criterion of a temperature phase transition in the region of a spin-glass phase diagram, where replicas are not present, 
this is so called the Nishimori line $e^{\beta}=\frac{1-p}{p}$ \cite{Nishimori_book, Nishimori}, ($p$ is concentration of ferromagnetic bonds). 

In these terms, the percolation transition temperature and a percolation cluster may be found \cite{0438, Yamaguchi} in frames of $Z(N)$.

Afterward, the future numerical estimations may be compared with the
mean field approximation results for investigated spin glasses
systems \cite{7}.

\textit{Acknowledgements.} The author thanks Professors B. I. Ostrovskii, M. I.
Polikarpov, and M. N. Chernodub for helpful discussions, and also
the Neutron Spectroscopy Department of Research Institute for Solid
State Physics and Optics of the Hungarian Academy of Sciences and
Laboratoire de Physique des Solides of Universit\'{e} de Paris-SUD
for hospitality.

\end{document}